\documentclass[aps,prd,eqsecnum,preprint,tightenlines,nofootinbib,showpacs]{revtex4-1}
\usepackage{graphicx,amsmath,latexsym}

\def\bea{\begin{eqnarray}}
\def\eea{\end{eqnarray}}
\def\be{\begin{equation}}
\def\ee{\end{equation}}

\begin{document}

\title{Dynamical model for longitudinal wave functions \\
in light-front holographic QCD
}

\author{Sophia S. Chabysheva}
\author{John R. Hiller}
\affiliation{Department of Physics \\
University of Minnesota-Duluth \\
Duluth, Minnesota 55812}

\date{\today}

\begin{abstract}
We construct a Schr\"odinger-like equation for the longitudinal
wave function of a meson in the valence $q\bar{q}$ sector,
based on the 't~Hooft model for large-$N$ two-dimensional QCD,
and combine this with the usual transverse equation from
light-front holographic QCD, to obtain a model for mesons
with massive quarks.  The computed wave functions are compared
with the wave function {\em ansatz} of Brodsky and De T\'eramond
and used to compute decay constants and parton distribution
functions.  The basis functions used to solve the longitudinal
equation may be useful for more general calculations of
meson states in QCD.
\end{abstract}
%
%
%

%

\maketitle

\section{Introduction}
\label{sec:Introduction}

Light-front holographic QCD~\cite{LFhQCD} exploits an approximate
AdS$_5$/QCD duality to obtain a Schr\"odinger-like equation for
the transverse wave functions of hadrons, for massless quarks.
Although a true QCD dual theory is unknown, the approximate
duality with AdS$_5$ can be obtained by altering the geometry
of AdS$_5$ at infrared scales corresponding to the QCD
confinement scale, $1/\Lambda_{\rm QCD}$.  The modifications
incorporate confinement, and, in the modified AdS$_5$ theory,
one obtains the point-like behavior of QCD partons and dimensional
counting rules~\cite{PolchinskiStrassler}.  The approximate
duality leads to a boost-independent light-front equation
for the valence state of a hadron~\cite{LFhQCD,formfactormatch,LFhQCD2}.
This equation provides a first-order approximation to the light-front
QCD eigenvalue problem for hadrons in the valence Fock sector,
but only for massless quarks.

From a modeling perspective, light-front holographic QCD generates
an effective potential for valence quarks from the choice
of the warping of the AdS$_5$ dual, rather than modeling the
effective potential directly.  This leads to a light-front equation
for the transverse wave function for massless quarks.  The longitudinal
wave function is determined by correspondence with a form factor
in the AdS$_5$ dual~\cite{formfactormatch}, instead of a dynamical
equation. The soft-wall model~\cite{softwall}
in particular admits analytic solutions for the spectrum
and the transverse wave functions, and, what is more,
yields a spectrum~\cite{Gershtein,Vega,Ebert,Branz,Kelley}
consistent with linear Regge trajectories.
However, because the duality relies on the approximate
conformal limit of zero-mass quarks, the transverse
and longitudinal wave functions do not include dependence on
quark masses.

For realistic calculations, we need to make an extension to
massive quarks.  In light-front coordinates, the mass is
associated with the longitudinal part of the kinetic energy,
as given in (\ref{eq:invariantmass}), not the transverse.
Thus the introduction of massive quarks requires specification
of either the longitudinal wave function or a dynamical
equation for it.  Brodsky and De T\'eramond have provided
an {\em ansatz} for the former~\cite{ansatz}; the purpose
of this paper is to consider the latter possibility
and compare with the Brodsky--De T\'eramond {\em ansatz}.

The Brodsky--De T\'eramond {\em ansatz} extends
the transverse momentum dependence to include the full
invariant mass.  For meson states, where the transverse
wave function can be a gaussian, this extends an exponentiation
of $k_\perp^2/x(1-x)$, where $\vec k_\perp$ is the transverse
momentum and $x$ is the longitudinal momentum fraction,
to $k_\perp^2/x(1-x)+\mu_1^2/x+\mu_2^2/(1-x)$.  To be consistent
with the zero-mass limit, the quark masses $\mu_1$ and $\mu_2$
are current quark masses, the parameters in the QCD Lagrangian.
Constituent quark masses are nonzero even when the current
quark masses are zero.

An alternative, which we consider here,
is to assume separation of variables for
the meson wave function and then provide a light-front equation 
for the longitudinal part that includes quark masses and
a model potential.  The longitudinal kinetic term will be just 
$m_1^2/x+m_2^2/(1-x)$; here, to be consistent with
nonrelativistic quark models, the quark masses $m_1$ and $m_2$
are constituent masses.  The potential term should be confining
and should yield longitudinal wave functions consistent with the
dual AdS$_5$/QCD form-factor analysis, in the zero-current-mass limit.
A potential model that achieves this, and is directly related
to QCD, is the 't~Hooft model obtained in the
large-$N$ limit of two-dimensional QCD~\cite{tHooft}.  It is
just such a instantaneous gluon-exchange potential that appears in
four-dimensional QCD in light-cone gauge.  The structure of our
quark model, with its longitudinal/transverse separation,
is then a direct analog of transverse-lattice
QCD~\cite{TransLatticeBP,TransLatticeBPR}, where the 't~Hooft
model provides the longitudinal connection between transverse
lattice planes.

In the remaining sections, we explore this potential-model
approach.\footnote{This is
quite different from a holographic description of the 't~Hooft
model itself~\protect\cite{KatzOkui}, where one considers
an AdS$_3$ dual to two-dimensional QCD.}  In Sec.~\ref{sec:motivation}
we provide some background details and the motivations for
our choice of longitudinal potential.
The details of the model and its solution are discussed in Sec.~\ref{sec:model};
sample applications are illustrated in Sec.~\ref{sec:sample}.
A summary and some additional remarks are included in Sec.~\ref{sec:summary}.
Appendices contain details of the conventions for light-front coordinates,
of the dual form-factor analysis, and of the numerical
solution for the model.

\section{Motivation} \label{sec:motivation}

To define our model, we begin from an effective
light-front Schr\"odinger equation for the quark-antiquark
wave function $\psi(x,\vec k_\perp)$ of a meson
\be \label{eq:fullLFSE}
\left[ \frac{\mu_1^2}{x}+\frac{\mu_2^2}{1-x}+\frac{k_\perp^2}{x(1-x)}
    +\widetilde U\right]\psi=M^2\psi,
\ee
where the first three terms are the kinematic invariant mass,
as discussed in \ref{sec:LFcoords}, and
$\widetilde U$ is an effective potential.  A transverse
Fourier transform to a relative coordinate $\vec b_\perp$ yields
\be
\left[ \frac{\mu_1^2}{x}+\frac{\mu_2^2}{1-x}-\frac{1}{x(1-x)}\nabla_\perp^2
    +\widetilde U\right]\psi=M^2\psi,
\ee
with 
$\nabla_\perp^2=\frac{\partial^2}{\partial b_\perp^2}
  +\frac{1}{b_\perp}\frac{\partial}{\partial b_\perp}
  +\frac{1}{b_\perp^2}\frac{\partial^2}{\partial \varphi^2}$
the transverse Laplacian and $\varphi$ the polar angle.  The choice
of a new coordinate $\zeta\equiv\sqrt{x(1-x)}b_\perp$ is then
convenient.  Combined with the factorization
\be
\psi=e^{iL\varphi}X(x)\phi(\zeta)/\sqrt{2\pi\zeta}
\ee
and the natural
assumption that $\widetilde U$ conserves the angular momentum component
$L_z$, the light-front equation (\ref{eq:fullLFSE}) reduces to~\cite{LFhQCD}
\be \label{eq:reducedLFSE}
\left[ \frac{\mu_1^2}{x}+\frac{\mu_2^2}{1-x}-\frac{\partial^2}{\partial\zeta^2}
-\frac{1-4L^2}{4\zeta^2}+\widetilde U\right]X(x)\phi(\zeta)=M^2X(x)\phi(\zeta).
\ee
For zero-mass quarks, $\widetilde U$ becomes just $U(\zeta)$, a function 
of $\zeta$ only, and the longitudinal wave function $X$ is no longer
determined by Eq.~(\ref{eq:reducedLFSE}), which
leaves a one-dimensional equation for the transverse wave function $\phi(\zeta)$
\be \label{eq:transverse}
\left[ -\frac{d^2}{d\zeta^2}
-\frac{1-4L^2}{4\zeta^2}+ U(\zeta)\right]\phi(\zeta)=M^2\phi(\zeta).
\ee

The assumed duality with AdS$_5$ can suggest models for $U$, through
a correspondence between the transverse Schr\"odinger equation (\ref{eq:transverse})
and the equation of motion for a spin-$J$ field in AdS$_5$~\cite{LFhQCD}.
Confinement is introduced by a dilaton profile $\widetilde\phi(z)$, where $z$
is the holographic coordinate of AdS$_5$.  With the identification
of $z$ with $\zeta$~\cite{LFhQCD}, the corresponding effective
potential is~\cite{effU}
\be
U(\zeta)=\frac12\widetilde\phi''(\zeta)+\frac14\widetilde\phi'(\zeta)^2
+\frac{2J-3}{2\zeta}\widetilde\phi'(\zeta).
\ee
For the soft-wall model~\cite{softwall}, the dilaton profile is
$\widetilde\phi(z)=e^{\pm \kappa^2 z^2}$, with $\kappa$ a parameter,
and the effective potential reduces to an oscillator potential
\be
U(\zeta)=\kappa^4\zeta^2+2\kappa^2(J-1).
\ee
For this potential, the spectrum of masses is $M^2=4\kappa^2\left(n+(J+L)/2\right)$,
with $n$ the radial quantum number, and the transverse wave functions are the
two-dimensional oscillator eigenfunctions.  The spectrum of the model provides
for a linear Regge trajectory and a good fit to light meson masses~\cite{spectrum}.

The longitudinal wave function $X$ remains unspecified.  It can, however,
be constrained by the duality in an analysis of the meson form factor~\cite{LFhQCD}.
As summarized in \ref{sec:dualFF}, this leads to the conclusion that
$X(x)=\sqrt{x(1-x)}$ for massless quarks.

For massive quarks, something needs to be assumed, beyond the
AdS$_5$ correspondence.  One approach, as already mentioned,
is the {\em ansatz}
by Brodsky and De T\'eramond~\cite{ansatz}.  Since the transverse
wave functions are harmonic oscillator eigenfunctions, the
ground state is a simple Gaussian $e^{-\kappa^2\zeta^2/2}=e^{-\kappa^2 x(1-x)b_\perp^2/2}$.
The transform to transverse momentum is, of course, also
a Gaussian, $\frac{4\pi^2}{\kappa^2}\frac{1}{x(1-x)}e^{-k_\perp^2/(2\kappa^2 x(1-x))}$.
The {\em ansatz} replaces $k_\perp^2/(x(1-x))$ with
$k_\perp^2/x(1-x)+\mu_1^2/x+\mu_2^2/(1-x)$.  The form of $X(x)$ is
then
\be \label{eq:ansatz}
X_{\rm BdT}(x)=N_{\rm BdT}\sqrt{x(1-x)}e^{-(\mu_1^2/x+\mu_2^2/(1-x))/2\kappa^2},
\ee
with $N_{\rm BdT}$ a normalization factor.  The $\sqrt{x(1-x)}$ form is
recovered in the zero-mass limit.

In our approach, we start from the full light-front Schr\"odinger
equation (\ref{eq:fullLFSE}) and replace the effective potential
$\widetilde U$ by $U(\zeta)+U_\parallel$ and the current masses
$\mu_i$ by constituent masses $m_i$.  The potential $U_\parallel$
is an integral operator which acts on functions of momentum
fraction $x$.  The combination of this
longitudinal potential and the change in mass is meant to 
represent the longitudinal effects of the original (unknown)
effective potential $\widetilde U$; the specific choice of constituent
masses is driven by consistency with nonrelativistic quark models.
The equation then factorizes into the transverse equation
\be
\left[ -\frac{d^2}{d\zeta^2}
-\frac{1-4L^2}{4\zeta^2}+ U(\zeta)\right]\phi(\zeta)=(M^2-M_\parallel^2)\phi(\zeta),
\ee
which differs from (\ref{eq:transverse}) only by the separation
constant $M_\parallel^2$, and the longitudinal equation
\be  \label{eq:longitudinal}
\left[\frac{m_1^2}{x} +\frac{m_2^2}{1-x}+U_\parallel\right]X(x)=M_\parallel^2 X(x).
\ee
The effective longitudinal potential $U_\parallel$ can be adjusted to
make $M_\parallel^2$ equal to zero for the ground state; this allows
the fit of $M^2$ to the meson mass spectrum to remain unaffected.  Also,
because the transverse spectrum is a good fit, we consider only
the ground state for the longitudinal equation.

Our choice for the longitudinal potential $U_\parallel$ is the 't~Hooft 
model~\cite{tHooft} obtained in the large-$N$ limit of two-dimensional
QCD.  The selection is motivated
by two factors.  First, it is the natural choice for a confining
potential in one spatial dimension, particularly for modeling the
longitudinal part of three-dimensional QCD.  When QCD is quantized in
light-cone gauge, such a potential appears automatically as
an instantaneous Coulomb-like interaction between quark currents.
For this reason, it is also part of the longitudinal interaction
included in transverse lattice gauge 
theory~\cite{TransLatticeBP,TransLatticeBPR}, where
fields on transverse nodes and links are coupled longitudinally
by a continuum model, a transverse/longitudinal separation
not unlike the situation for light-front holographic QCD.

Second, there exists a nearly exact analytic solution
for the ground state, which can be improved easily with 
numerical calculations~\cite{Bergknoff,MaHiller,MoPerry}
and which can be arranged to be consistent with the expected
$X(x)$ in the zero-current-mass limit.  As is
known from the work of 't~Hooft~\cite{tHooft} and Bergknoff~\cite{Bergknoff},
the approximate analytic solution is of the form $x^{\beta_1}(1-x)^{\beta_2}$,
with $\beta_i$ determined by the quark masses and the longitudinal
coupling.  For equal constituent masses, the coupling can be
adjusted to obtain the desired $\beta_i=1/2$ for zero current
masses.  We also use this condition to fix the value
of the longitudinal coupling.

\section{The model}
\label{sec:model}

With the longitudinal potential taken from the 't~Hooft model~\cite{tHooft},
the longitudinal equation (\ref{eq:longitudinal}) becomes
\be \label{eq:longitudinaleqn}
\left[\frac{m_1^2}{x}+\frac{m_2^2}{1-x}\right]X(x)
  +\frac{g^2}{\pi}{\cal P}\int dy \frac{X(x)-X(y)}{(x-y)^2}-C X(x)=M_\parallel^2 X(x),
\ee
with ${\cal P}$ indicating the principal value and $C$ a constant.
Because the transverse equation
already introduces enough quantum numbers, we consider only the
ground state of the longitudinal equation; any additional quantum
number associated with longitudinal excitations would represent 
double counting.  Also, since the light-meson spectrum is already
represented by the transverse equation, the constant $C$ is used
to set $M_\parallel$ to zero.  The net effect is that the additional
longitudinal equation is only for determination of the longitudinal
wave function and has nothing to say about the spectrum.

As discussed in the previous section,
the ground-state wave function $X(x)$ is well approximated by
the form $x^{\beta_1}(1-x)^{\beta_2}$.
From the dual form-factor analysis~\cite{formfactormatch}, the light-meson
wave function should have this form with $\beta_1=\beta_2=1/2$.
Analysis of the endpoint behavior for the solution of
(\ref{eq:longitudinaleqn}) shows that $\beta_i$ should satisfy the
transcendental equation~\cite{tHooft,Bergknoff}
\be
\frac{m_i^2 \pi}{g^2}-1+\pi\beta_i \cot\pi\beta_i=0.
\ee
If we take the up and down quark masses, $m_u$ and $m_d$, to be equal,
the square-root behavior is obtained if $g^2/\pi=m_u^2$,
consistent with $\cot\pi/2=0$.  This
fixes the value of the coupling constant.  Although the model
could be more flexible if $g$ were flavor dependent, we do
not consider this.

The exact solution for the wave function is not analytic.
However, a numerical solution, as presented in \ref{sec:numerical},
is straightforward.

For the ground state, the complete wave function is given by
a normalized product of the longitudinal wave function $X$
and a transverse Gaussian
\be
\psi(x,\zeta)=N X(x) e^{-\kappa^2\zeta^2/2},
\ee
or, in terms of the transverse coordinate $b_\perp$,
\be
\psi(x,b_\perp)=N X(x) e^{-\kappa^2 x(1-x)b_\perp^2/2}.
\ee
The factor $N$ is fixed by the normalization
\be
P_{q\bar{q}}=\int_0^1 dx \int_0^\infty \,db_\perp^2 \pi|\psi(x,b_\perp)|^2,
\ee
where $P_{q\bar{q}}$ is the probability of the quark-antiquark
valence state.  If $X(x)$ is separately normalized such that
\be
\int_0^1 dx \frac{|X(x)|^2}{x(1-x)}=1,
\ee
then $N=\frac{\kappa}{\pi}\sqrt{P_{q\bar{q}}}$.  

The wave functions can be used to compute decay constants and parton
distributions.  A decay constant is given by~\cite{decayconstant}
\be
f_M=2\sqrt{6}\int_0^1 dx \int_0^\infty \frac{dk_\perp^2}{16\pi^2} \psi(x,k_\perp).
\ee
As discussed in \cite{Vega} and shown in \cite{pdf},
the parton distribution $f(x)$ is given by
\be
f(x)=\frac{\kappa^2}{16\pi^2}x(1-x) \eta^2(x),
\ee
if the wave function takes the form
\be
\psi(x,k_\perp)=\eta(x) e^{-k_\perp^2/(2 \kappa^2 x(1-x))}.
\ee
Applying this to our model, we obtain
\be
f(x)=P_{q\bar{q}}\frac{X^2(x)}{x(1-x)}.
\ee
For the {\em ansatz}, we have
\be
f_{\rm BdT}(x)=N_{\rm BdT}^2 P_{q\bar{q}}  e^{-(\mu_1^2/x+\mu_2^2/(1-x))/\kappa^2},
\ee
with the normalization of the {\em ansatz} given by
\be
N_{\rm BdT}=\left[\int_0^1 dx e^{-(\mu_1^2/x+\mu_2^2/(1-x))/\kappa^2}\right]^{-1/2}.
\ee
%

\section{Sample calculations}
\label{sec:sample}

To see the implications of our model, we compare the
form of the longitudinal wave function with the
{\em ansatz} by Brodsky and De T\'eramond~\cite{ansatz},
both directly and through the computation of 
decay constants and parton distribution functions,
for the pion, kaon, and J/$\Psi$.  Where
parameter values are needed, we use the current-quark
parameterization
of Vega {\em et al}.~\cite{Vega} with no additional
fits or adjustments.  The parameter values are listed
in Table~\ref{tab:parameters}.

\begin{table}[ht]
\caption{\label{tab:parameters} Meson parameters and decay constants.
All dimensionful parameters are in units of MeV.  Results are compared
between our model, which uses constituent-quark masses,
and the {\em ansatz} of Brodsky and De T\'eramond~\protect\cite{ansatz},
with current-quark masses.  Parameter and experimental values
are from Vega {\em et al}.~\protect\cite{Vega} and
the Particle Data Group~\protect\cite{PDG}.
}
\begin{center}
\begin{tabular}{cccccccccc}
\hline \hline
 & \multicolumn{2}{c}{model} &  \multicolumn{2}{c}{\em ansatz} &
  &   &  \multicolumn{3}{c}{decay constant} \\
meson & $m_1$ & $m_2$ & $\mu_1$ & $\mu_2$ & $P_{q\bar{q}}$ & $\kappa$ & model & {\em ansatz} & exper. \\
\hline
pion & 330 & 330 & 4 & 4 & 0.204 & 951 & 131 & 132 & 130 \\
kaon & 330 & 500 & 4 & 101 & 1 & 524 & 160 & 162 & 156 \\
J/$\Psi$ & 1500 & 1500 & 1270 & 
  1270 & 1 & 894 & 267 & 238 & 278 \\
\hline \hline
\end{tabular}
\end{center}
\end{table}

\begin{figure}
\vspace{0.2in}
\begin{center}
\begin{tabular}{cc}
\multicolumn{2}{c}{\includegraphics[width=6cm]{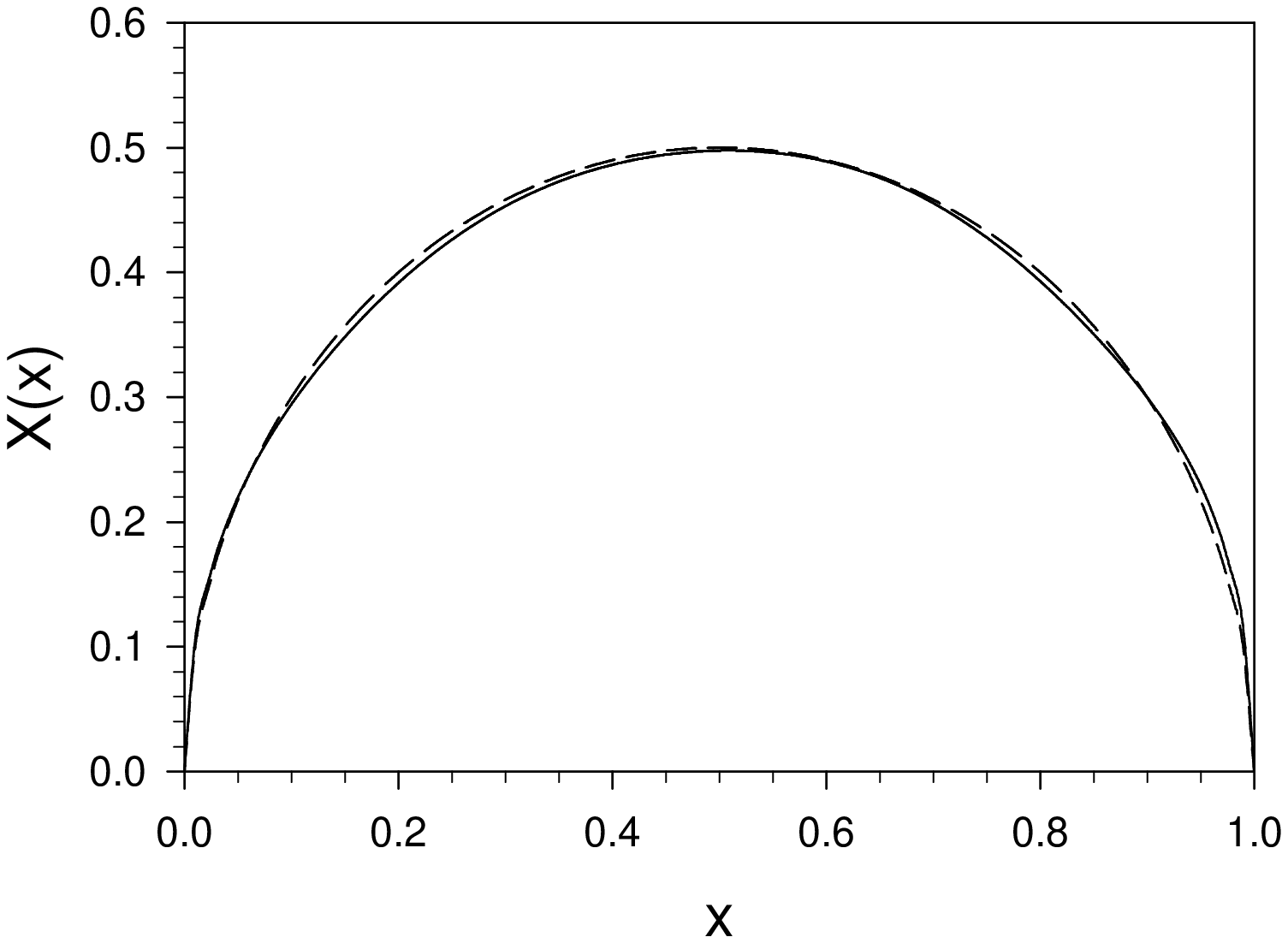}} \\
\multicolumn{2}{c}{(a)} \\
\\ \\
\includegraphics[width=6cm]{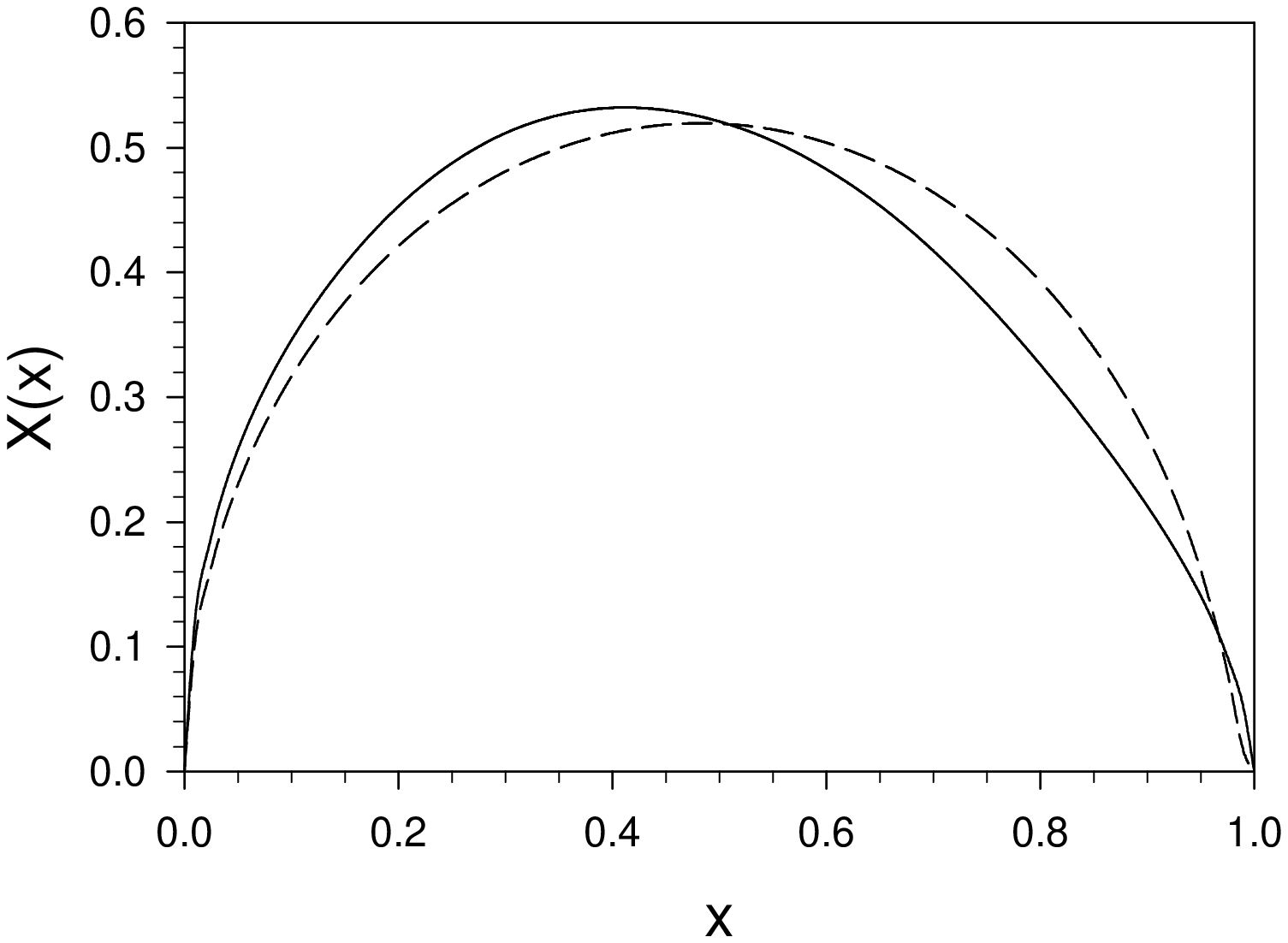} &
\includegraphics[width=6cm]{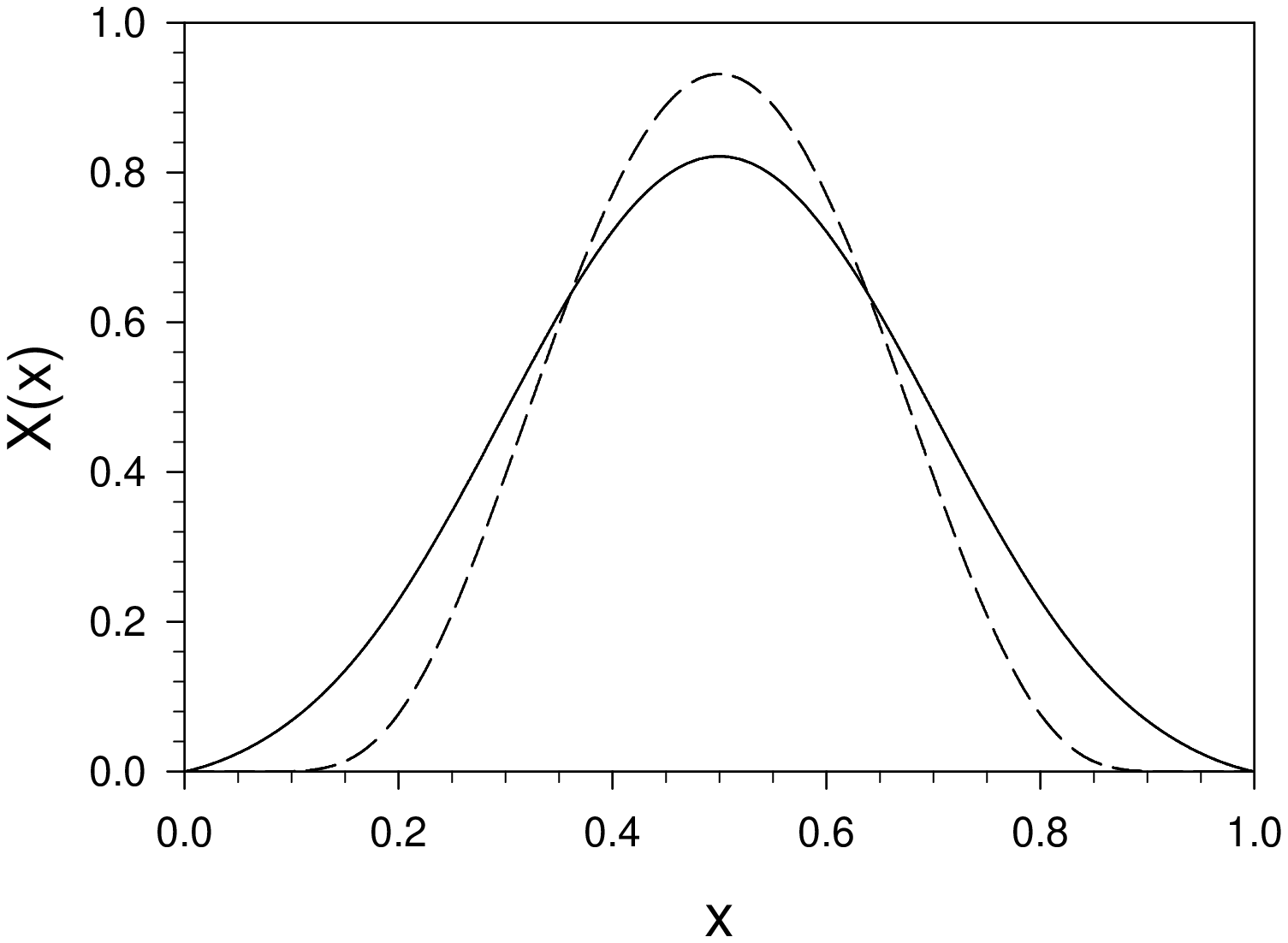} \\
(b) & (c)
\end{tabular}
\end{center}
\caption{\label{fig:longwf}Longitudinal wave functions $X(x)$ for
the (a) pion, (b) kaon, and (c) J/$\Psi$.  The solid lines are
wave functions from our model; the dashed lines show the {\em ansatz}
by Brodsky and De~T\'eramond~\protect\cite{ansatz}.
}
\end{figure}

Figure~\ref{fig:longwf} compares the {\em ansatz} with the $X(x)$ 
computed in our model.  For the pion, the two wave functions 
are essentially the same, since both involve only tiny
variations from the wave function $\sqrt{x(1-x)}$ for
quarks with zero current mass.

The results for decay constants are included in Table~\ref{tab:parameters}.
The values for the pion and kaon are consistent and in agreement with experiment.
Our model value for the J/$\Psi$ is significantly closer to experiment
than the value obtained from the longitudinal {\em ansatz}.

\begin{figure}[ht]
\vspace{0.2in}
\begin{center}
\begin{tabular}{cc}
\multicolumn{2}{c}{\includegraphics[width=6cm]{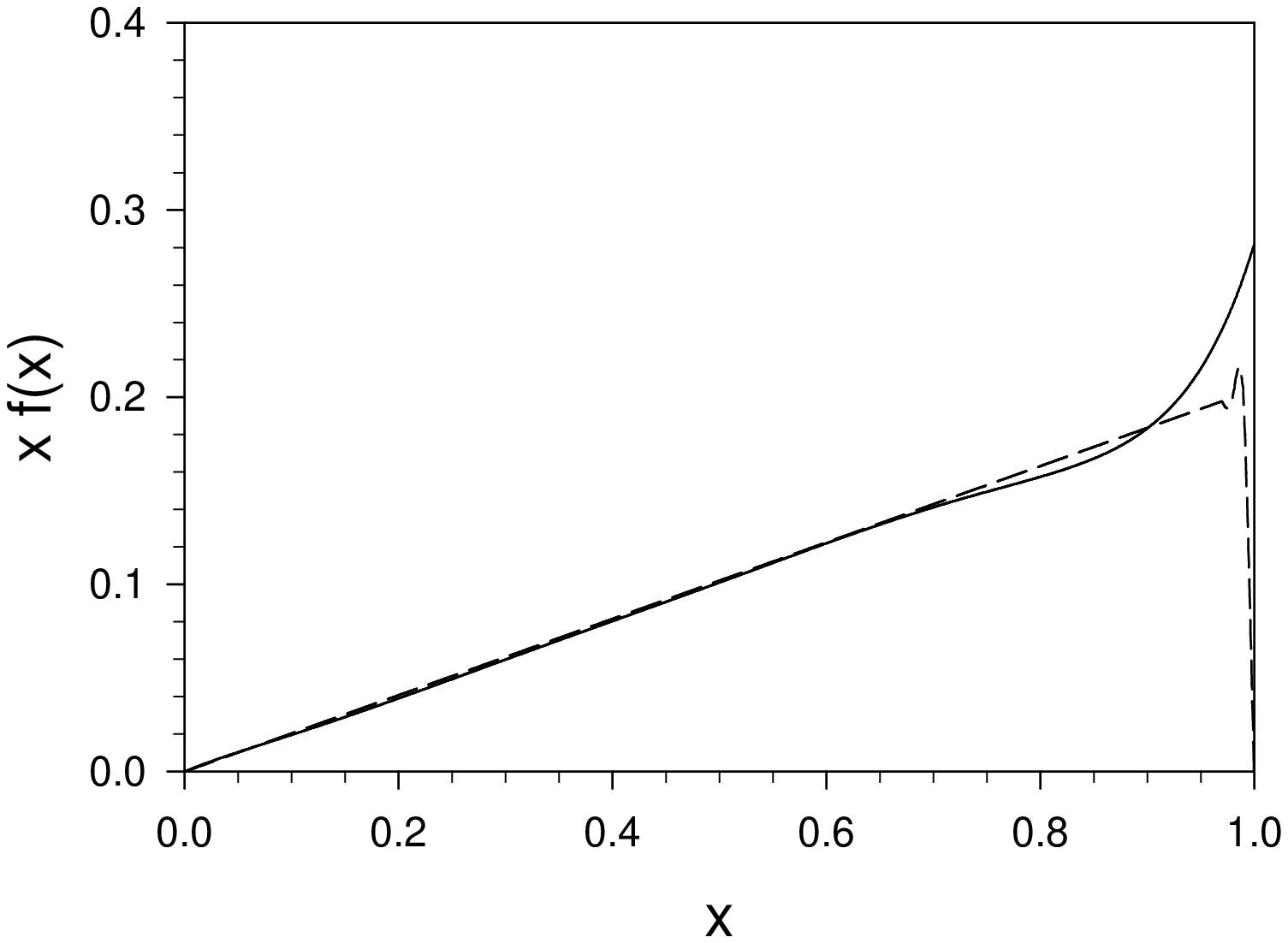}} \\
\multicolumn{2}{c}{(a)} \\
\\ \\
\includegraphics[width=6cm]{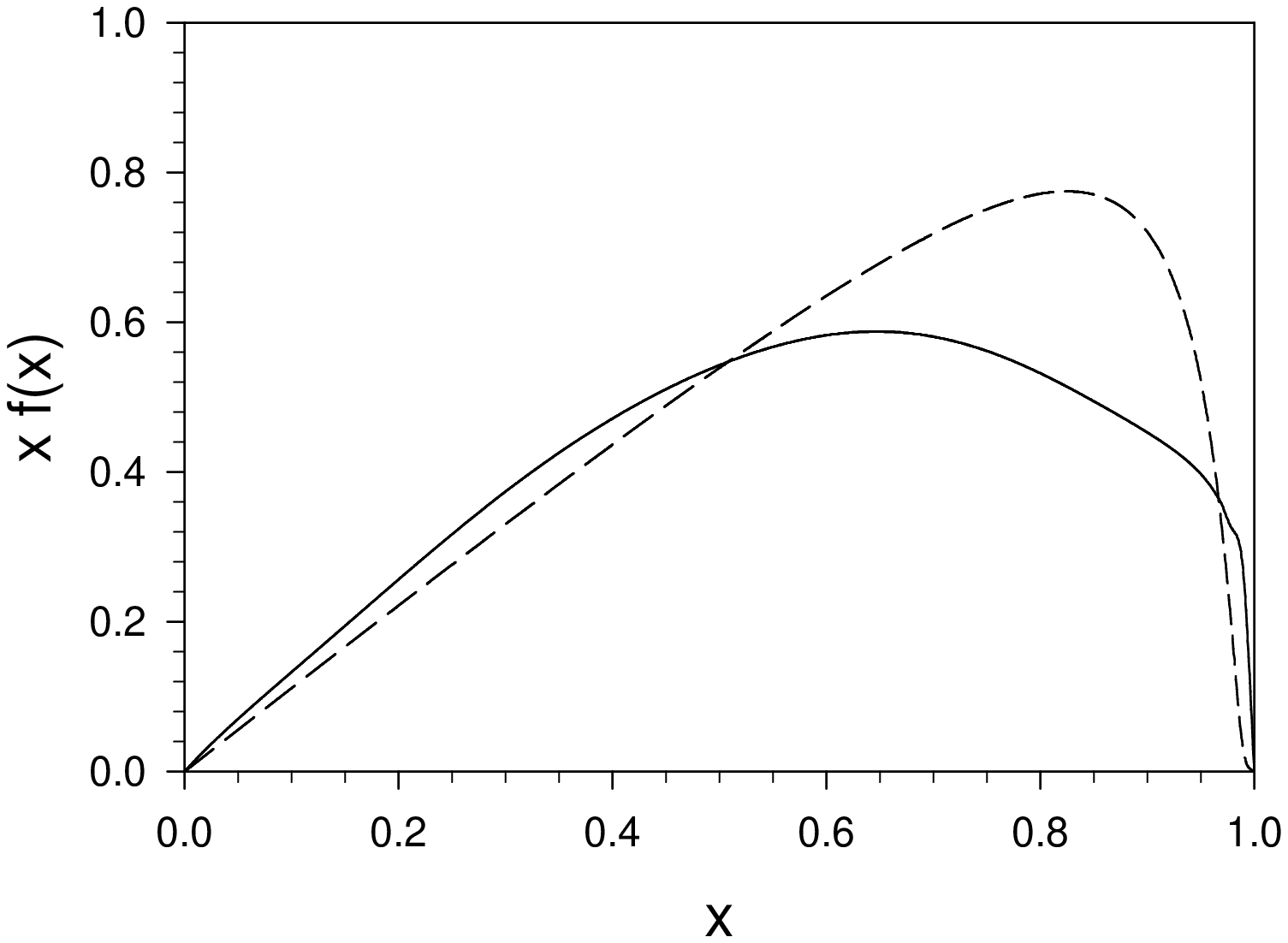} &
\includegraphics[width=6cm]{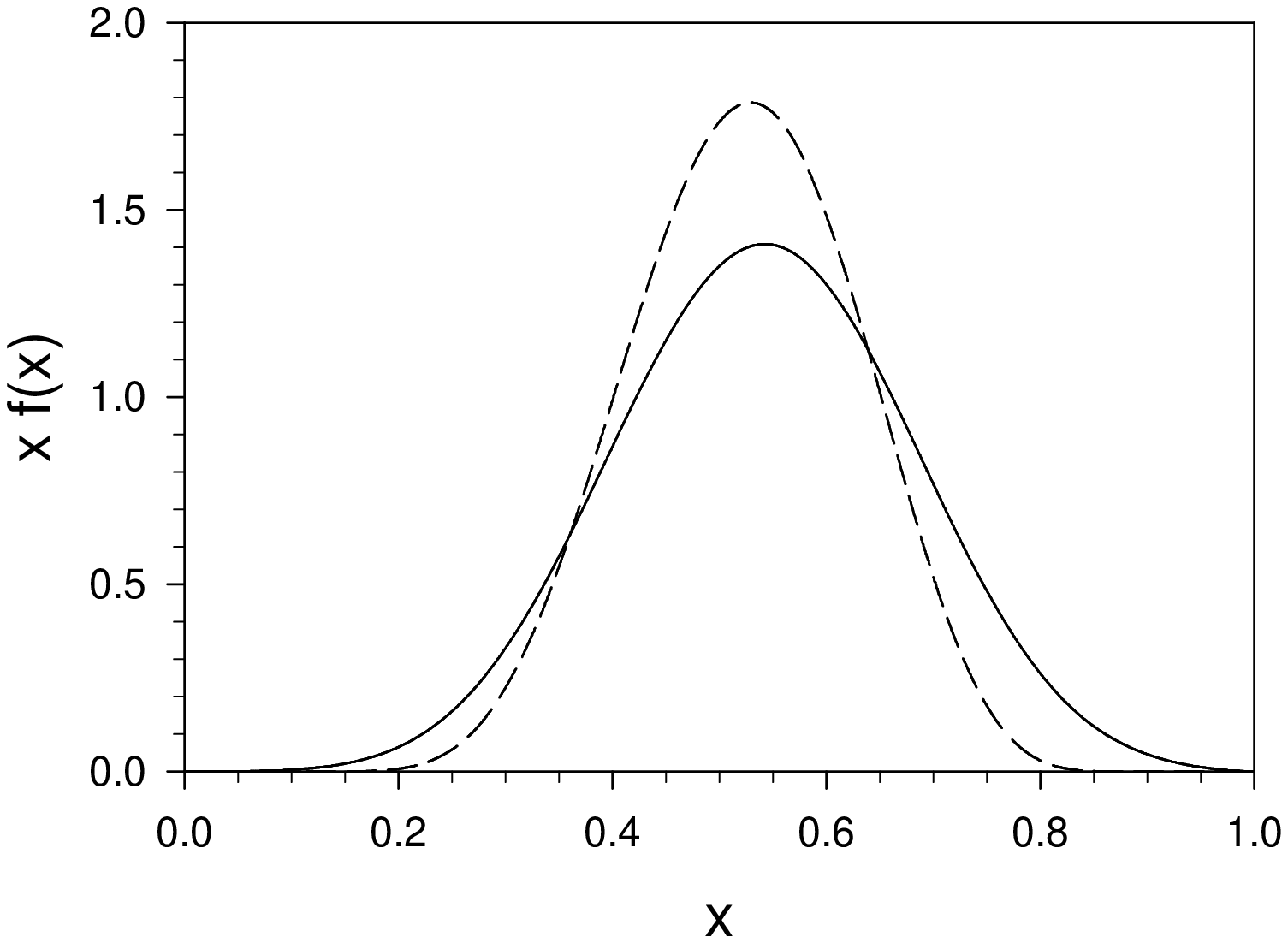} \\
(b) & (c)
\end{tabular}
\end{center}
\caption{\label{fig:pdf} Same as Fig.~\protect\ref{fig:longwf}
but for parton distributions $f(x)$ multiplied by $x$.
}
\end{figure}

The parton distributions are plotted for comparison in Figure~\ref{fig:pdf}.
The rough similarity of the wave functions translates into similar parton
distributions.

\section{Concluding remarks}
\label{sec:summary}

We have constructed and solved a relativistic light-front equation
for the longitudinal wave functions of mesons with
massive quarks, to be used in tandem with the transverse
equation of light-front holographic QCD.  Comparisons with
the {\em ansatz}~\cite{ansatz} (\ref{eq:ansatz}) show 
that for lighter mesons, the longitudinal
wave functions are quite similar.  However, for the J/$\Psi$,
there is a notable difference, which translates
into a better estimate of the decay constant, as listed
in Table~\ref{tab:parameters}.  Wave functions for
the J/$\Psi$, and also the pion and kaon, are shown
in Fig.~\ref{fig:longwf}, and parton distributions
are shown in Fig.~\ref{fig:pdf}.  The similarity of
the longitudinal wave functions provides the {\em ansatz}
with a connection to the fundamental interactions of QCD.

Perhaps the most broadly useful outcome of this exercise
is to illustrate that there is a convenient set of basis
functions for longitudinal wave functions of meson
valence states.  These are the $f_n$ defined in
(\ref{eq:basis}), with the parameters $\beta_i$
to be optimized as needed for a given application.
One could, of course, use the
eigenfunctions of the 't Hooft model, but these do
not have an analytic form and would add an extra
layer of complication to any calculation.

The importance of the choice of basis functions
is in the rate of convergence as the basis is
expanded.  For the alternative of discrete
light-cone quantization, it is known that 
convergence can be much slower~\cite{vandesande}.
Thus, these basis functions may prove useful
for calculations, such as those described in \cite{Vary},
which use the transverse light-front holographic eigenfunctions.

\acknowledgments
This work was supported in part by the Department of Energy
through Contract No.\ DE-FG02-98ER41087.
We thank G.F. de~T\'eramond and S.J. Brodsky for helpful
comments.

\appendix

\section{Light-front coordinates}  \label{sec:LFcoords}

Our conventions for light-front coordinates~\cite{Dirac,DLCQreviews}
are as follows.  We define light-front time $x^+=t+z$
and the longitudinal light-front spatial coordinate $x^-=t-z$.
The transverse coordinates are collected as $\vec x_\perp=(x,y)$.
The corresponding light-front energy and momentum are
$p^-=E-p_z$, $p^+=E+p_z$, and $\vec p_\perp=(p_x,p_y)$.
From the invariant mass relation $m^2=p^2=p^+p^--\vec p_\perp^2$,
we obtain $p^-=(m^2+p_\perp^2)/p^+$ for an on-shell particle.

For a system of particles, with total momentum $(P^+,\vec P_\perp)$,
the longitudinal momentum fraction $x_i=p_i^+/P^+$ and
relative transverse momentum $\vec k_{\perp i}=\vec p_{\perp i}-x_i\vec P_\perp$,
for the ith particle, are boost invariant and therefore a convenient
choice for independent variables in the description of the system.
They sum to one and zero, respectively:
\be
\sum_i x_i=1,\;\;\;\; \sum_i\vec k_{\perp i}=0.
\ee
The invariant mass for the system is
\be \label{eq:invariantmass}
P^+\sum_i p_i^--\vec P_\perp^2
   =\sum_i \frac{m_i^2+(x_i\vec P_\perp+\vec k_{\perp i})^2}{p_i^+/P^+}-\vec P_\perp^2
   =\sum_i \frac{m_i^2+\vec k_{\perp i}^2}{x_i}.
\ee
For a two-particle system, the internal variables reduce to 
$x=x_1$, $x_2=1-x$, $\vec k_\perp=\vec k_{\perp 1}$, and $\vec k_{\perp 2}=-\vec k_\perp$,
and the invariant mass becomes
\be
\frac{m_1^2+\vec k_{\perp 1}^2}{x_1}+\frac{m_2^2+\vec k_{\perp 2}^2}{x_2}
=\frac{m_1^2}{x}+\frac{m_2^2}{1-x}+\frac{\vec k_\perp^2}{x(1-x)}.
\ee
%

\section{Dual form-factor analysis}  \label{sec:dualFF}

In the Drell--Yan--West frame~\cite{DrellYan}, the form factor for momentum
transfer $q^2$ can be written
in terms of Fock-state wave functions $\psi_n$ as~\cite{BrodskyDrell}
\be \label{eq:formfactor}
F(q^2)=\sum_n\int [dx_i] [d^2k_{\perp i}]\sum_j e_j
    \psi_n^*(x_i,\vec k_{\perp i}^{\,\prime})\psi_n(x_i,\vec k_{\perp i}),
\ee
where $n$ denotes the Fock sector, $e_j$ the charge of the $j$th quark,
\be
[dx_i]\equiv \prod_{i=1}^n \int dx_i \delta(1-\sum_j x_j),
\ee
\be
[d^2k_{\perp i}]\equiv \left(\prod_{i=1}^n 
   \int\frac{d^2k_{\perp i}}{16\pi^3}\right) 16 \pi^3 \delta(\sum_j \vec k_{\perp j}),
\ee
and, with $j$ the index of the quark that absorbed the photon,
\be
\vec k_{\perp i}^{\,\prime}=\left\{\begin{array}{ll} 
             \vec k_{\perp i}+(1-x)\vec q_\perp, & i=j \\
             \vec k_{\perp i}-x\vec q_\perp, & i\neq j. \end{array}\right.
\ee
Substitution of the wave function as a transverse Fourier transform in impact space,
\be
\psi_n(x_i,\vec k_{\perp i})=(4\pi)^{(n-1)/2}
    \prod_{i=1}^{n-1} \int d^2b_{\perp i} 
        e^{i\sum_{j=1}^{n-1}\vec b_{\perp j}\cdot\vec k_{\perp j}}
        \widetilde\psi_n(x_i,\vec b_{\perp i}),
\ee
converts the expression (\ref{eq:formfactor}) for the form factor to
\be
F(q^2)=\sum_n \prod_{j=1}^{n-1}\int dx_j d^2b_{\perp j}
    e^{i\vec q_\perp\cdot\sum_{j=1}^{n-1}x_j\vec b_{\perp j}}
       |\widetilde\psi_n(x_i,\vec b_{\perp i})|^2.
\ee
For the quark-antiquark valence sector alone, with the wave function
given by $\widetilde\psi_2=e^{iL\varphi}X(x)\phi(\zeta)/\sqrt{2\pi\zeta}$,
this reduces to
\be
F(q^2)=\int \frac{dx\,|X(x)|^2}{x(1-x)} \int d\zeta 
     J_0\left(\zeta q_\perp\sqrt{x/(1-x)}\right)|\phi(\zeta)|^2,
\ee
with $J_0$ the Bessel function of order zero.
This is to be compared with the form computed in
AdS$_5$~\cite{PolchinskiStrassler}
\be
F(q^2)=\int dx \int d\zeta 
     J_0\left(\zeta q_\perp\sqrt{x/(1-x)}\right)|\phi(\zeta)|^2.
\ee
Thus, the conclusion~\cite{LFhQCD} that $X(x)=\sqrt{x(1-x)}$,
when the quarks have zero current mass.

\section{Numerical solution}  \label{sec:numerical}

We solve (\ref{eq:longitudinaleqn}) numerically by expanding the
wave function $X$ in terms of orthonormal basis functions $f_n$,
chosen to include the analytic approximation explicitly. 
As discussed by Mo and Perry~\cite{MoPerry}, these basis
functions are
\be  \label{eq:basis}
f_n(x)=N_n x^{\beta_1}(1-x)^{\beta_2}P_n^{(2\beta_2,2\beta_1)}(2x-1),
\ee
with $P_n^{(2\beta_2,2\beta_1)}$ the Jacobi polynomial of order $n$.
The normalization factor $N_n$ is given by~\cite{AbramowitzStegun}
\be
N_n=\sqrt{(2n+2\beta_1+2\beta_2)
   \frac{n! \Gamma(n+2\beta_1+2\beta_2+1)}{\Gamma(n+2\beta_1+1)\Gamma(n+2\beta_2+1)}}.
\ee
The solution is then represented as
\be
X(x)=\sum_n c_n f_n(x).
\ee
For equal-mass cases, the longitudinal equation obeys an $x\leftrightarrow(1-x)$
symmetry, and only the even-$n$ terms will contribute.  In general, we find
that only a few terms are needed; the $n=0$ term, for which the Jacobi polynomial
is constant and $f_0\propto x^{\beta_1}(1-x)^{\beta_2}$, represents 90\% or
more of the probability.

The expansion coefficients $c_n$ are obtained by diagonalizing the
longitudinal equation in the $f_n$ basis.  The matrix representation is
\be
\left( \frac{m_1^2}{m_u^2}A_1+\frac{m_2^2}{m_u^2}A_2+B\right)\vec c=\xi\vec c,
\ee
with $\xi\equiv C/m_u^2$ and matrices $A_1$, $A_2$, and $B$ defined by
\bea
(A_1)_{nm}&=&\int_0^1 \frac{dx}{x} f_n(x) f_m(x),\;\;
(A_2)_{nm}=\int_0^1 \frac{dx}{1-x} f_n(x) f_m(x), \\
B_{nm}&=&\int_0^1 dx \,{\cal P}\!\!\int_0^1 dy f_n(x)\frac{f_m(x)-f_m(y)}{(x-y)^2}.
\eea
Following 't~Hooft~\cite{tHooft},
the matrix representation is made explicitly symmetric by rewriting
the potential term as
\be
B_{nm}=\frac12\int_0^1 dx \int_0^1 dy 
   \frac{f_n(x)-f_n(y)}{x-y}\frac{f_m(x)-f_m(y)}{x-y}.
\ee
In addition to the explicit symmetry, which simplifies the
matrix diagonalization, this rearrangement also
resolves the principal value prescription.  The matrices are
small because the number of terms needed in the expansion
are few; the diagonalization is then straightforward.



\begin{thebibliography}{}

\bibitem{LFhQCD} S.J. Brodsky and G.F. de T\'eramond,
Phys.\ Rev.\ Lett.\ {\bf 96} (2006), 201601.

\bibitem{PolchinskiStrassler} J. Polchinski and M.J. Strassler,
Phys.\ Rev.\ Lett.\ {\bf 88} (2002), 031601;
JHEP {\bf 05} (2003), 012.

\bibitem{formfactormatch} S.J. Brodsky and G.F. de T\'eramond,
Phys.\ Rev.\ D {\bf 77} (2008), 056007.

\bibitem{LFhQCD2} S.J. Brodsky and G.F. de T\'eramond,
Phys.\ Lett.\ B {\bf 582} (2004), 211;
J. Erlich, E. Katz, D.T. Son, and M.A. Stephanov,
Phys.\ Rev.\ Lett.\ {\bf 95}, 261602 (2005);
Z. Abidin and C.E. Carlson, Phys.\ Rev.\ D {\bf 77} (2008), 095007;
S.J. Brodsky and G.F. de T\'eramond, Phys.\ Rev.\  {\bf 78} (2008), 025032;
G.F. de T\'eramond and S.J. Brodsky, Phys.\ Rev.\ Lett.\ {\bf 102} (2009), 081601.

\bibitem{softwall} A. Karch, E. Katz, D.T. Son, and M.A. Stephanov,
Phys.\ Rev.\ D {\bf 74} (2006), 015005.

\bibitem{Gershtein} S.S. Gershtein, A.K. Likhoded, and A.V. Luchinsky,
Phys.\ Rev.\ D {\bf 74} (2006), 016002.
   
\bibitem{Vega} A. Vega, I. Schmidt, T. Branz, T. Gutsche, and V.E. Lyubovitskij,
Phys.\ Rev.\ {\bf 80} (2009), 055014.

\bibitem{Ebert} D. Ebert, R.N. Faustov, and V.O. Galkin,
Eur.\ Phys.\ J.\ C {\bf 66} (2010), 197.

\bibitem{Branz} T. Branz, T. Gutsche, V.E. Lyubovitskij, I. Schmidt, and A. Vega,
Phys.\ Rev.\ D {\bf 82} (2010), 074022;
T. Gutsche, V.E. Lyubovitskij, I. Schmidt, and A. Vega,
Phys.\ Rev.\ D {\bf 85} (2012), 076003.

\bibitem{Kelley} T.M. Kelley, S.P. Bartz, and J. Kapusta,
Phys.\ Rev.\ D {\bf 83} (2011), 016002.

\bibitem{ansatz} S.J. Brodsky and G.F. de T\'eramond, arXiv:0802.0514.

\bibitem{tHooft} G. 't Hooft,
Nucl.\ Phys.\ B {\bf 75} (1974), 461.

\bibitem{TransLatticeBP} W.A. Bardeen and R.B. Pearson,
Phys.\ Rev.\ D {\bf 14} (1976), 547.

\bibitem{TransLatticeBPR}
W.A. Bardeen, R.B. Pearson, and E. Rabinovici,
Phys.\ Rev.\ D {\bf 21} (1980), 1037.

\bibitem{KatzOkui} E. Katz and T. Okui,
JHEP {\bf 901} (2009), 013.

\bibitem{effU} G.F. de T\'eramond and S.J. Brodsky,
AIP Conf.\ Proc.\ {\bf 1296} (2010), 128.

\bibitem{spectrum} G.F. de T\'eramond and S.J. Brodsky,
arXiv:1203.4025.

\bibitem{Bergknoff} H. Bergknoff,
Nucl.\ Phys.\ B {\bf 122} (1977), 215.

\bibitem{MaHiller} Y. Ma and J.R. Hiller,
J. Comput.\ Phys.\ {\bf 82} (1989), 229.

\bibitem{MoPerry} Y. Mo and R.J. Perry,
J. Comput.\ Phys.\ {\bf 108} (1993), 159.

\bibitem{decayconstant} S.J. Brodsky, T. Huang, and G.P. Lepage,
{\em Proceedings of the Banff Summer Institute on Particles
and Fields 2, Banff, Alberta, 1981}, edited by A.Z. Capri and A.N. Kamal
(Plenum, New York, 1983), p.~143;
G.P. Lepage, S.J. Brodsky, T. Huang, and P.B. Mackenzie,
{\em ibid.}, p.~83;
T. Huang, 
AIP Conf.\ Proc.\ {\bf 68} (1980), 1000.

\bibitem{pdf} A.V. Radyushkin,
Phys.\ Rev.\ D {\bf 58} (1998), 114008.

\bibitem{PDG} K. Nakamura {et al}.(Particle Data Group),
J. Phys.\ G {\bf 37} (2010), 075021.

\bibitem{vandesande} B. van de Sande,
Phys.\ Rev.\ D {\bf 54} (1996), 6347.

\bibitem{Vary} J.P. Vary {\em et al}.,
Phys.\ Rev.\ C {\bf 81} (2010), 035205.

\bibitem{Dirac} {P.A.M. Dirac, 
Rev.\ Mod.\ Phys.\ {\bf 21} (1949), 392.}

\bibitem{DLCQreviews} For reviews, see
   M. Burkardt, Adv.\ Nucl.\ Phys.\ {\bf 23} (2002), 1;
   S.J. Brodsky, H.-C. Pauli, and S.S. Pinsky, 
   Phys.\ Rep.\ {\bf 301} (1998), 299.
   
\bibitem{DrellYan} S.D. Drell, D.J. Levy, and T.M. Yan,
Phys.\ Rev.\ Lett.\ {\bf 22} (1969), 744;
G.B. West, Phys.\ Rev.\ Lett.\ {\bf 24} (1970), 1206.

\bibitem{BrodskyDrell} S.J. Brodsky and S.D. Drell,
Phys.\ Rev.\ D {\bf 22} (1980), 2236.

\bibitem{AbramowitzStegun} M. Abramowitz and I.A. Stegun (eds.),
{\em Handbook of Mathematical Functions} (Dover, New York, 1965).

\end{thebibliography}
\end{document}